\documentclass[english,twocolumn]{revtex4}
\usepackage[latin1]{inputenc}
\usepackage{amsmath}
\usepackage{graphicx}
\usepackage{amssymb}

\makeatletter
\usepackage{revsymb}

\renewcommand{\vec}[1]{\mathbf{#1}}

\renewcommand*{\sectionmark}[1]{} 
\renewcommand*{\subsectionmark}[1]{} 

\usepackage{babel}
\makeatother
\begin{document}

\newcommand{\ket}[1]{|#1\rangle}

\newcommand{\bra}[1]{\langle#1|}

\newcommand{\braopket}[3]{\left\langle #1\left|#2\right|#3\right\rangle }

\newcommand{\bsigma}{\boldsymbol{\sigma}}

\newcommand{\brho}{\boldsymbol{\rho}}

\newcommand{\bzeta}{\boldsymbol{\zeta}}

\newcommand{\bvarepsilon}{\boldsymbol{\varepsilon}}

\newcommand{\ShermanBmAvg}{\gamma}

\newcommand{\cSOLeff}{c^{*}}

\newcommand{\fExpCoeffZero}{C_{k}}

\newcommand{\fExpCoeffLinearSO}{D_{k}}

\newcommand{\SOcoupling}{\lambda}

\newcommand{\Egap}{E_{0}}

\newcommand{\DeltaSO}{\Delta_{0}}

\newcommand{\polAHE}{\vec{P}_{\mathrm{AH}}}

\newcommand{\JAHE}{\vec{J}_{\mathrm{AH}}}

\newcommand{\jCCC}[3]{j_{#1,\,#3}^{#2}}

\newcommand{\jSH}{\vec{j}}

\newcommand{\jSHc}{j}

\newcommand{\jSHSJ}{\jSH_{\,\mathrm{SJ}}}

\newcommand{\sigmaSHSS}{\sigma_{\mathrm{SS}}^{\mathrm{SH}}}

\newcommand{\unitVector}[1]{\hat{\vec{#1}}}

\newcommand{\crossSecSpin}{d\negmedspace\!\stackrel{\leftrightarrow}{\sigma}\negmedspace\!}

\global\emergencystretch = 0.2\hsize

\title{Theory of Spin Hall conductivity in $n$-doped GaAs }

\author{Hans-Andreas Engel, Bertrand I. Halperin, Emmanuel I. Rashba}

\affiliation{Department of Physics, Harvard University, Cambridge, Massachusetts
02138 }

\author{}

\begin{abstract}
We develop a theory of extrinsic spin currents in semiconductors,
resulting from spin-orbit coupling at charged scatterers, which leads
to skew scattering and side jump contributions to the spin Hall conductance.
Applying the theory to bulk $n$-GaAs, without any free parameters,
we find spin currents that are in reasonable agreement with recent
experiments by Kato \emph{et al.} {[}Science \textbf{306,} 1910 (2004){]}.
\end{abstract}


\maketitle

Generating and manipulating non-equilibrium spin magnetization by
electric fields is one of the most desirable goals of semiconductor
spintronics, because electric fields have potentialities for accessing
individual spins at nanometer scale. Spin-orbit (SO) coupling is a
mechanism for achieving this goal. It has the prominent consequence
of the spin-Hall effect (SHE), where an electric-current can induce
a transverse spin current and a non-equilibrium spin accumulation
near sample boundaries. Recent observations of this effect are important
experimental achievements \cite{KatoSpinHall,Wunderlich05}. Theoretically,
two different mechanisms of SHE were proposed. The \textit{extrinsic}
mechanism \cite{DP71,Hirsch99,Zhang00} is based on spin-dependent
scattering of electrons by impurities and is mostly due to Mott skew
scattering \cite{MottMassey}. An \textit{intrinsic} mechanism has
also been proposed, which is related to the concept of dissipationless
spin currents in a perfect crystal \cite{Murakami03,Sinova04}. 

The theory of spin transport in media with SO coupling is rather intricate
and includes all problems inherent in the theory of anomalous Hall
effect (AHE), which has a long history; for reviews see \cite{Nozieres73,CrepieuxBrunoAHE}.
Also, the precise definition of spin currents is still under dispute
because of the problems stemming from spin nonconservation in media
with SO coupling. If spin-relaxation rates are small, a spin current
with non-zero divergence can lead to spin accumulations, which are
experimentally observable quantities \cite{SHList,Mish04,fnKatoModel}.
However, because spin-currents are even with respect to time inversion,
they do not necessarily vanish in thermodynamic equilibrium, hence
their relation to spin transport and spin accumulation is far from
obvious \cite{R03}. 

These problems inherent in the spin-transport theory make identification
of physical mechanisms underlying the SHE observed in Refs.~\onlinecite{KatoSpinHall} and \onlinecite{Wunderlich05}
rather challenging. On the one hand, Wunderlich \textit{et al.} \cite{Wunderlich05}
observed a strong SHE in two-dimensional (2D) layers of $p$-GaAs
and ascribed it to the \emph{intrinsic} effect because of the large
magnitude of the effect and large splitting of the energy spectrum
typical of heavy holes. On the other hand, Kato \textit{et al.}~\cite{KatoSpinHall}
attribute their measurement of SHE in three-dimensional $n$-GaAs
layers (2$\mu$m thick) to the \emph{extrinsic} mechanism. We believe
that this is indeed the case as we explain in this work. Although
an intrinsic spin Hall effect, driven by the $k^{3}$-Dresselhaus
SO coupling \cite{Dresselhaus55}, could give rise to spin accumulation
in this system, as proposed in Ref.~\onlinecite{BernevigZhangIntrinsic3D},
its estimated size, when impurity scattering is taken into account,
is an order of magnitude smaller than the observations. Further, because
of the large sample thickness, a specifically 2D mechanism of spin
accumulation proposed recently be Usaj and Balseiro \cite{Usaj04}
and Nikoli\'{c} \textit{et al.} \cite{Nikolic2DBoundary} and relying
on the properties of near-edge states cannot play a role in the geometry
of Ref.~\onlinecite{KatoSpinHall}. 

In the following, we develop a theory of extrinsic spin currents in
a 3D electron system. It results from intrinsic SO coupling in the
bulk crystal that produces a SO contribution to the impurity potential.
(This effect can occur even when the host crystal is inversion symmetric,
so that spin is conserved outside the radii of the impurity potentials.)
We find that impurity scattering and SO interaction in $n$-GaAs are
strong enough to support spin-currents that are in reasonably good
agreement with findings by Kato \textit{et al.}~\cite{KatoSpinHall}
without using any adjustable parameters. Although the sign of our
result is opposite to that mentioned in Ref.~\onlinecite{KatoSpinHall},
there remains some uncertainty about the absolute sign in the experiments
\cite{KatoAwschalomPrivate}. Since our result contains two contributions
of opposite sign, and we have employed several approximations, we
would believe the extrinsic mechanism to be relevant even if it turns
out that the difference in the sign of the effect persists.

\begin{figure}
\begin{center}\includegraphics[%
  height=30mm,
  keepaspectratio]{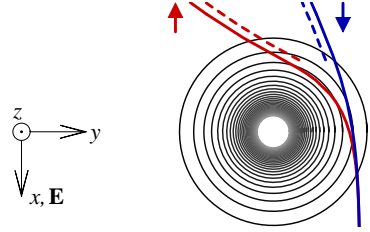}\end{center}

\caption{Spin-dependent scattering at an attractive impurity. We show the
classical trajectories (solid lines), for a screened Coulomb potential
and for strongly exaggerated spin-orbit coupling with $\SOcoupling>0$
and with quantization axis $\unitVector{z}$. The skew-scattering
current results from different scattering angles for spin-$\uparrow$
and spin-$\downarrow$ and leads to a positive spin Hall conductivity,
$\sigmaSHSS=-\jSHc_{\mathrm{SS},y}^{z}/E_{x}>0$. Further, we show
the horizontal displacement due to the side jump effect (dashed lines),
contributing to the spin current with opposite sign.}
\end{figure}

We consider an electron Hamiltonian of form \begin{equation}
H=\frac{\hbar^{2}k^{2}}{2m^{*}}+V(\vec{r})+\SOcoupling\:\bsigma\cdot\left(\vec{k}\times\nabla V\right),\label{eq:H}\end{equation}
where $V(\vec{r})$ is an electron potential energy that varies slowly
on the scale of the host lattice constant. In vacuum, the last term
of Eq.~(\ref{eq:H}) results from relativistic corrections in the
Pauli equation and is known as Thomas term, with $\SOcoupling=-\hbar^{2}/4m_{0}^{2}c^{2}\approx-3.7\times10^{-6}\:\mbox{\AA}{}^{2}$,
vacuum electron mass $m_{0}$, and velocity of light $c$. In direct
gap cubic semiconductors like GaAs, a SO interaction of the same form
develops in the framework of the $\vec{k}\cdot\vec{p}$ model due
to the coupling of a $s$-type conductance band to $p$-type valence
bands. This coupling is well established and, for asymmetric 2D systems,
it leads to the Rashba term. For conduction band electrons in the
$8\times8$ Kane model, one finds $\SOcoupling=\left(P^{2}/3\right)[1/\Egap^{2}-1/\left(\Egap+\DeltaSO\right)^{2}]$
in third-order perturbation theory, with gap $E_{0}$, SO splitting
$\Delta_{0}$ between the $J=\frac{3}{2}$ and $J=\frac{1}{2}$ hole
bands, and a properly normalized interband matrix element $P$ of
the momentum \cite{Winkler}. All these parameters are large (i.e.,
of atomic scale) since they result from the strong crystal potential.
For example, for GaAs one finds $\SOcoupling=5.3\:\mbox{\AA}{}^{2}$.
Thus, the SO coupling in $n$-GaAs is by \textit{\emph{six orders
of magnitude stronger}} than in vacuum and has the \emph{opposite
sign}. This enhancement of SO coupling is critical for developing
large extrinsic spin currents. 

In the 3D problem that we consider, SO coupling comes only through
the potential energy $V(\vec{r})$ of Eq.~(\ref{eq:H}), that can
be either electron energy in an impurity center or its energy $-e(\vec{E}\cdot\vec{r})$
in a driving electric field, where $e$ is the charge of an electron,
i.e., $e<0$. We ignore here the SO $k^{3}$ (Dresselhaus) correction
to the electron energy in pure GaAs, which is small and is absent
in the $8\times8$ Kane model.

We now analyze the effect of the SO coupling in Eq.~(\ref{eq:H})
on the scattering at impurities, which leads to the extrinsic spin
Hall effect. This comprises two contributions, one resulting from
the skew-scattering at the impurities \cite{MottMassey} and the other
from the shift of the scattered wave packet \cite{BergerSideJump}
(known as side jump contribution). These contributions are shown schematically
in Fig.~1. 

The skew scattering contribution is calculated in the lowest order
both in SO interaction and in $1/k_{\mathrm{F}}\ell$, with Fermi
momentum $k_{F}$ and mean free path $\ell$. To this end, we describe
the system by a spin-dependent Boltzmann equation and a distribution
function written as a $2\times2$ spin matrix $\hat{f}=\left[f_{0}(\vec{k})+\phi(\vec{k})\right]1\!\!1+\vec{f}(\vec{k})\cdot\bsigma$,
with equilibrium distribution function $f_{0}$.  In the following,
we suppress the identity matrix $1\!\!1$. The collision integral
on the right hand side of the Boltzmann equation has the form \begin{equation}
-\left(\frac{\partial\hat{f}(\vec{k})}{\partial t}\right)_{\mathrm{coll}}=n_{i}\sum_{\vec{k}';\:\, k'=k}\,\frac{\hbar k}{m^{*}}\,\frac{\crossSecSpin}{d\Omega}\left[\hat{f}(\vec{k})-\hat{f}(\vec{k}')\right],\label{eqp:collTerm}\end{equation}
 where $n_{i}$ is the impurity density. The scattering cross section
$\crossSecSpin/d\Omega$ is spin-dependent and mixes spin components
of the incoming flux. In the general case, this spin dependence is
rather complex \cite{MotzRMP}. However, it simplifies essentially
when we expand Eq.~(\ref{eqp:collTerm}) in SO coupling and neglect
the cross terms containing spin-dependent contributions of both $\crossSecSpin/d\Omega$
and $\hat{f}$. For a central symmetric impurity potential, we can
then write \begin{align}
 & \frac{\crossSecSpin}{d\Omega}\left[\hat{f}(\vec{k})-\hat{f}(\vec{k}')\right]=I(\vartheta)\left[\hat{f}(\vec{k})-\hat{f}(\vec{k}')\right]\nonumber \\
 & \qquad+I(\vartheta)S(\vartheta)\bsigma\cdot\vec{n}\left[\phi(\vec{k})-\phi(\vec{k}')\right],\label{sigterms}\end{align}
 where $\vartheta=\vartheta_{\vec{k}\vec{k'}}$ is the angle between
$\vec{k}'$ and $\vec{k}$, and $\vec{n}=\vec{k}'\times\vec{k}/|\vec{k}'\times\vec{k}|$
is the unit vector normal to the scattering plane. The coefficient
$I(\vartheta)$ is the spin-independent part of the scattering cross
section, while $S(\vartheta)$ is the so-called Sherman function \cite{MottMassey,MotzRMP,VoskoboynikovPRB},
which measures the polarization of outgoing particles scattered into
direction $\vec{k}$ from an unpolarized incoming beam of momentum
$\vec{k}'$. 

To lowest order in the electric field, the left hand side of the Boltzmann
equation equals $e\vec{E}\cdot\frac{\partial f_{0}}{\partial\vec{p}}=(e\hbar/m^{*})(\vec{E}\cdot\vec{k})\,\frac{\partial f_{0}}{\partial\varepsilon}$,
where the isotropy of $f_{0}(\vec{k})$ was used. The Boltzmann equation
may then be solved with the following ansatz. First, the usual spin-independent
term is set to $\phi(\vec{k})=\vec{k}\cdot\vec{E}\:\fExpCoeffZero$
{[}typically, $C_{k}=-\left(e\hbar\tau/m^{*}\right)\frac{\partial f_{0}}{\partial\varepsilon}$
with transport lifetime $\tau${]}. Second, we use $\vec{f}(\vec{k})=\left(\vec{E}\times\vec{k}\right)\fExpCoeffLinearSO$
for the spin polarization due to SO interaction. This structure is
motivated by the physics of Mott scattering, namely, that the spin
polarization is perpendicular to the scattering plane defined by incoming
electrons that drift in the direction of $-\vec{E}$ and are scattered
into $\vec{k}$. Here, $C_{k}$ and $D_{k}$ are spherically symmetric
functions of $\vec{k}$. 

With this ansatz, we can evaluate the collision integrals on the right
hand side of the Boltzmann equation. Integrating over the direction
of $\vec{k}'$, $d\Omega\left(\vec{k}'\right)=d\varphi\, d\vartheta\sin\vartheta$,
where $\varphi$ is the the azimuth of $\vec{k}'$ in the plane perpendicular
to $\vec{k}$, and suppressing an overall factor of $n_{i}\hbar k/m^{*}$
in Eq.~(\ref{eqp:collTerm}), we obtain (see App.~\ref{app})\begin{align}
 & \int d\Omega(\vec{k}')\: I\left(\vartheta\right)\,\left(\vec{k}-\vec{k}'\right)\cdot\left(\vec{E}\,\fExpCoeffZero+\bsigma\times\vec{E}\, D_{k}\right)\nonumber \\
 & =\left[\vec{k}\cdot\vec{E}\,\fExpCoeffZero+\vec{k}\cdot(\bsigma\times\vec{E})D_{k}\right]\:\int d\Omega\: I\left(\vartheta\right)\,\left(1-\cos\vartheta\right),\label{eq:Icoll}\end{align}
 where the integral on the right is proportional to the inverse transport
time $\tau^{-1}$, and \begin{align}
 & \int d\Omega(\vec{k}')\: I\left(\vartheta\right)S\left(\vartheta\right)\left(\bsigma\cdot\vec{n}\right)\left(\vec{k}\cdot\vec{E}-\vec{k}'\cdot\vec{E}\right)\fExpCoeffZero\nonumber \\
 & =-\frac{1}{2}\,\vec{k}\cdot(\bsigma\times\vec{E})\, C_{k}\,\int d\Omega\: I\left(\vartheta\right)S\left(\vartheta\right)\sin\vartheta.\label{eq:IScoll}\end{align}

Since the left hand side of the Boltzmann equation only depends on
the component of $\vec{k}$ along the electrical field, this must
also be the case on the right hand side. Thus, the second term in
Eq.~(\ref{eq:Icoll}) must cancel with Eq.~(\ref{eq:IScoll}), determining
$D_{k}=\frac{1}{2}\ShermanBmAvg_{k}C_{k}$. We defined the transport
skewness\begin{equation}
\ShermanBmAvg_{k}=\frac{\int d\Omega\: I\left(\vartheta\right)S\left(\vartheta\right)\sin\vartheta}{\int d\Omega\: I\left(\vartheta\right)\,\left(1-\cos\vartheta\right)},\label{eq:GammaSkew}\end{equation}
 which describes the effect of skew-scattering on the distribution
function and depends on the structure of the scattering center and
on the energy of the scattered particle. Therefore, our ansatz is
self-consistent, and the solution of the Boltzmann equation is\begin{equation}
\hat{f}(\vec{k})=f_{0}(k)+\vec{k}\cdot\left[\vec{E}+\frac{\ShermanBmAvg_{k}}{2}\left(\bsigma\times\vec{E}\right)\right]\, C_{k},\label{eq:f}\end{equation}
i.e., the components of $\vec{f}(\vec{k})$ are $f_{\mu}(\vec{k})=\frac{\ShermanBmAvg_{k}}{2}\left(\vec{E}\times\vec{k}\right)_{\mu}C_{k}$. 

Now we calculate the contribution of skew-scattering to the spin current,
$\jSH_{\,\mathrm{SS}}^{\,\mu}=n\left\langle \sigma_{\mu}\,\vec{v}_{0}\right\rangle $,
with density $n$ and with $\vec{v}_{0}=\hbar\,\vec{k}/m^{*}$ (SO
contributions to the velocity are analyzed below). We obtain \begin{equation}
\jCCC{\mathrm{SS}}{\mu}{\kappa}=\mathrm{Tr}\,\sigma_{\mu}\!\int\frac{d^{3}k}{\left(2\pi\right)^{3}}\,\frac{\hbar\, k_{\kappa}}{m^{*}}\:\hat{f}(\vec{k})=\frac{\ShermanBmAvg}{2e}\,\varepsilon^{\kappa\mu\nu}\left(\vec{J}_{0}\right)_{\nu},\label{eq:SS}\end{equation}
where $\vec{J}_{0}=2e\int d^{3}k\,(2\pi)^{-3}\left(\hbar\,\vec{k}/m^{*}\right)\:\vec{k}\cdot\vec{E}\, C_{k}$
is the charge current in the absence of SO coupling, and summation
over $\nu$ is implied. Assuming low temperatures, we need to evaluate
$\ShermanBmAvg_{k}$ only near the Fermi energy $E_{\mathrm{F}}$
and we defined $\ShermanBmAvg\equiv\ShermanBmAvg_{k_{\mathrm{F}}}$
\cite{Skewness2D}. If there are different species of impurity potentials,
the weighted average of the corresponding transport skewnesses $\ShermanBmAvg$,
Eq.~(\ref{eq:GammaSkew}), should be taken. Note that repulsive impurities
generally lead to the opposite sign of $\ShermanBmAvg$. This can
result in a partial suppression of spin Hall currents due to impurity
compensation (where some acceptors are still present in an $n$-doped
material and can localize electrons, leading to repulsive scatterers).

Next we evaluate the contribution of the side jump \cite{BergerSideJump}
to spin currents, where the wave function is laterally displaced during
the scattering event. (This displacement does not modify the scattering
angle measured at large distances, i.e., it does not affect the scattering
cross section.) Side-jump currents were analyzed in detail for the
anomalous Hall effect \cite{Nozieres73} and we now relate the AHE
to the SHE. In the AHE, a net polarization in combination with SO
interaction at impurities, leads to electrical Hall currents, even
in the absence of an external magnetic field. For the SHE, electrons
are unpolarized in equilibrium and we consider induced spin currents
$\jSH^{\,\mu}$. Because $\SOcoupling$ is small and spin relaxation
is of order $\SOcoupling^{2}$ \cite{Elliott,YafetSSP}, one can understand
the $\hat{\bvarepsilon}_{\mu}$-component of spin current as difference
in particle currents of two spin species with polarizations $\pm\hat{\bvarepsilon}_{\mu}$.
For non-interacting electrons, each of these species carries the anomalous
Hall current $\JAHE^{\uparrow,\downarrow}$ of a system with density
$n_{\mathrm{AH}}=\frac{1}{2}n$ and with spins fully aligned along
the $\pm\hat{\bvarepsilon}_{\mu}$ direction, and we can express the
spin Hall current as\begin{equation}
\jSH_{\,\mathrm{SH}}^{\,\mu}=e^{-1}\left(\JAHE^{\uparrow}-\JAHE^{\downarrow}\right).\label{eq:SHAH}\end{equation}

For the AHE, the side-jump contribution was found to be $\JAHE^{\mathrm{SJ},\,\uparrow}=-2\, n_{\mathrm{AH}}\SOcoupling\left(e^{2}/\hbar\right)\,\hat{\bvarepsilon}_{\mu}\times\vec{E}$
\cite{Nozieres73}. It results from SO corrections $\delta\dot{\vec{r}}$
to the velocity operator during impurity scattering. Nozi\`eres and
Lewiner \cite{Nozieres73} clarified that this anomalous velocity
$\delta\vec{\dot{r}}$ comprises two equal SO contributions. The first
is $\delta_{1}\dot{\vec{r}}=(i/\hbar)[H,\:\vec{r}]-\vec{v}_{0}=(\lambda/\hbar)(\nabla V\times\bsigma)$
and becomes $\delta_{1}\dot{\vec{r}}=\lambda(\bsigma\times\vec{\dot{k}})$
after the equation of motion, $\dot{\vec{k}}=-\nabla V/\hbar$, is
taken into account. The second originates from the correction to the
coordinate operator, the Yafet term $\delta\vec{r}_{\mathrm{SO}}=\SOcoupling(\bsigma\times\vec{k})$
\cite{YafetSSP}, and contributes as $\delta_{2}\dot{\vec{r}}=\delta_{1}\dot{\vec{r}}$.
Note that $\delta_{2}\dot{\vec{r}}$ leads to a factor of 2 which
is often ignored. Heuristically, we can now understand the current
$\JAHE^{\mathrm{SJ}}$ as follows. For scattering at an impurity with
momentum transfer $\delta\vec{k}$, the lateral displacement it $\delta\vec{r}=2\lambda(\bsigma\times\delta\vec{k})$.
The anomalous Hall current is recovered from $\JAHE^{\mathrm{SJ}}=en\,\delta\vec{r}/\tau$,
by using that the momentum dissipated per scattering event is $\hbar\delta\vec{k}=-e\vec{E}\tau$.
Using $\JAHE^{\mathrm{SJ}}$ \cite{Nozieres73} and Eq.~(\ref{eq:SHAH}),
we readily obtain the side jump contribution to the SHE, \begin{equation}
\jCCC{\mathrm{SJ}}{\mu}{\kappa}=-2n\SOcoupling\frac{e}{\hbar}\,\varepsilon^{\kappa\mu\nu}E_{\nu}.\label{eq:SJ}\end{equation}
 The sum $\jSH_{\,\mathrm{SS}}+\jSHSJ$ provides the total spin-Hall
current. 

Mathematically, one could consider a model where the electron charge
is cancelled by a uniform positive background, with only small fluctuations
in the potential $V$. In such a case, we could let $\tau$ become
arbitrarily large. Although the side-jump contribution to the spin
Hall conductance is independent of $\tau,$ the skew-scattering contribution,
given by Eq.~(\ref{eq:SS}), would grow with $\tau$. This growth
would be cut off, however, when $\tau$ becomes comparable to the
inverse of the SO splitting $\Delta$ due to the $k^{3}$ Dresselhaus
term. If $\gamma$ is small compared to $\Delta/E_{\mathrm{F}}$,
as would be the case if the potential fluctuations are sufficiently
small, then the skew-scattering contribution will be smaller than
the intrinsic contribution by a factor of order $\gamma\, E_{\mathrm{F}}/\Delta$.

We now evaluate the skewness $\ShermanBmAvg$ {[}Eq.~(\ref{eq:GammaSkew}){]}
for a screened attractive Coulomb potential. For this, we make use
of the long-established theory of single electron scattering by an
atom \cite{MotzRMP}. We rescale the parameters to make a connection
between the atomic Hamiltonian and the effective Hamiltonian {[}Eq.~(\ref{eq:H}){]}
with $V=-e^{-q_{s}r}\, e^{2}/\epsilon r$, effective mass $m^{*}$,
permittivity $\epsilon$, and screening length $1/q_{s}$. We match
$V$ be setting the atomic number to $Z=1/\epsilon$. Further, to
match the SO interaction, we define an {}``effective'' speed-of-light
$\cSOLeff$ such that $\SOcoupling=\hbar^{2}/4\left(m^{*}\cSOLeff\right)^{2}$.
Finally, when the sign of $\SOcoupling$ differs from its vacuum value,
we replace $S(\vartheta)$ by $-S(\vartheta)$ in Eq.~(\ref{eq:GammaSkew})
\cite{fnRescale}. For GaAs, $\epsilon=12.4$ and $m^{*}=0.0665\, m_{0}$,
thus $\cSOLeff\approx c/79$ and $\alpha^{*}Z\approx1/21$, with fine
structure constant $\alpha^{*}=e^{2}/\hbar c^{*}$. When evaluating
$I(\vartheta)$, screening of the long-range Coulomb interaction is
required, otherwise the integrals in Eq.~(\ref{eq:GammaSkew}) diverge,
e.g., $\tau\to0$. For exponential screening and in second order Born
approximation, $I(\vartheta)$ is given by the Dalitz formula \cite{MotzRMP}
(with the parameters used below and at large $\vartheta$, it agrees
well with the exact solution for the unscreened potential). We assume
Thomas-Fermi screening with inverse screening length $q_{s}=\sqrt{3e^{2}n/2\epsilon E_{\mathrm{F}}}.$
 The numerical evaluation below shows that the transport skewness
$\ShermanBmAvg$ depends only weakly on the screening, implying that
Eq.~(\ref{eq:SS}) is well-suited for evaluating the skew-scattering
contribution. {[}Because of low sensitivity of $\gamma$ to screening,
Eq.~(\ref{eq:SS}) is also in agreement with the common belief that
the skew-scattering contribution is proportional to $\tau$.{]} When
evaluating $S(\vartheta)$, we note that the main contribution comes
from large angles $\vartheta$ where the effects of screening are
negligible. We thus use for $S(\vartheta)$ the exact expression for
an unscreened potential \cite{MotzRMP,MottMassey}. Note that when
evaluating $S(\vartheta)$, the small parameter is $\left(Z\alpha^{*}\right)^{2}=4\left|\SOcoupling\right|/(a_{B}^{*})^{2}$,
with effective Bohr radius $a_{\mathrm{B}}^{*}=\hbar^{2}\epsilon/m^{*}e^{2}$,
so we expect $1/500$ for the order of magnitude of $\gamma$. 

We now estimate $\ShermanBmAvg$ for GaAs and for the electron density
$n=3\times10^{16}\:\mathrm{cm^{-3}}$ reported in Ref.~\onlinecite{KatoSpinHall},
i.e., $E_{\mathrm{F}}=5.3\:\mathrm{meV}$ and $q_{s}^{-1}\approx9\,\mathrm{nm}$.
As a test, we first evaluate the longitudinal conductivity using an
impurity density $n_{i}=n$ and the Drude formula, $\sigma_{xx}=e^{2}n\tau/m^{*}$
with $\tau^{-1}=n_{i}v_{\mathrm{F}}\int d\Omega\: I\left(\vartheta\right)\,\left(1-\cos\vartheta\right)$
and arrive at $\sigma_{xx}\approx1.8\times10^{3}\:\Omega^{-1}\mathrm{m}^{-1}$.
This is within 10\% of the experimentally observed conductivity at
low voltages; a surprisingly good agreement, given that the rather
small value $k_{\mathrm{F}}\ell\approx2$ restricts the accuracy of
the Boltzmann approach. Next we evaluate Eq.~(\ref{eq:GammaSkew})
and find $\ShermanBmAvg\approx1/900$. This value is rather stable
and changes by less than 30\% when $q_{s}$ or $E_{\mathrm{F}}$ increase
or decrease by a factor of two. 

Next, we estimate the spin Hall currents. The measurements were performed
at electrical fields $E\approx20\,\mathrm{mV\,\mu m}^{-1}$ where
the conductivity increased to $\sigma_{xx}\approx3\times10^{3}\:\Omega^{-1}\mathrm{m}^{-1}$
due to electron heating. We assume that $\ShermanBmAvg$ is not very
sensitive to these heating effects and we still use Eq.~(\ref{eq:SS})
but with the increased conductivity. For an electrical field $\vec{E}=\unitVector{x}\, E_{x}$,
we find both contributions to the spin Hall conductivity $\sigma^{\textrm{SH}}\equiv-\jSHc_{y}^{z}/E_{x}$,
namely $\sigmaSHSS=-(\ShermanBmAvg/2e)\sigma_{xx}\approx1.7\:\Omega^{-1}\mathrm{m}^{-1}/|e|$
and $\sigma_{\mathrm{SJ}}^{\mathrm{SH}}=2n\SOcoupling e/\hbar\approx-0.8\:\Omega^{-1}\mathrm{m}^{-1}/|e|$.
In total, we arrive at the extrinsic spin Hall conductivity $\sigma_{\mathrm{th.}}^{\mathrm{SH}}\approx0.9\:\Omega^{-1}\mathrm{m}^{-1}/|e|$
\cite{fnkFell}. The magnitude is within the error bars of the experimental
value of $\left|\sigma_{\mathrm{exp.}}^{\mathrm{SH}}\right|\approx0.5\:\Omega^{-1}\mathrm{m}^{-1}/|e|$
found from spin accumulation near the free edges of the specimen~\cite{KatoSpinHall,fnQualitative}.

Bernevig and Zhang calculated intrinsic spin currents due to $k^{3}$-Dresselhaus
interaction \cite{BernevigZhangIntrinsic3D}. In contrast to the case
of spin-orbit interactions linear in $\vec{k}$, where vertex corrections
lead to a cancellation of the spin current \cite{Inoue04,Mish04,Chalaev2D},
for the $k^{3}$ interaction this is not true; e.g., for $s$-wave
scatterers the vertex corrections vanish identically \cite{MishenkoPrivate}.
However, in the dirty limit $\tau\Delta/\hbar\sim10^{-2}$, where
$\Delta$ is the SO splitting $\Delta$ at the Fermi level and for
experimental parameters of Ref.~\onlinecite{KatoSpinHall}, intrinsic
spin currents were found to be small, only $\sigma_{\mathrm{D}}\approx0.02\:\Omega^{-1}\mathrm{m}^{-1}/|e|$
\cite{BernevigZhangIntrinsic3D}.

In conclusion, we solved the kinetic equation including skew-scattering
at impurities. We derived the contributions to the extrinsic spin
Hall effect and found reasonable agreement with experimental data.
We acknowledge discussions with D.D. Awschalom, B.A. Bernevig, E.
Demler, Y.K. Kato, D. Loss, A.H. MacDonald, E.G. Mishchenko, A. Stern,
R. Winkler, and S.-C. Zhang. This work was supported in part by NSF
grants DMR-02-33773, PHY-01-17795, the Harvard Center for Nanoscale
Systems, and DARPA.

\vfill

\appendix

\section{Angular Integrals}

\label{app} For completeness, we now show how the angular integrals
in Eqs.~(\ref{eq:Icoll}) and (\ref{eq:IScoll}) can be evaluated.
Without loss of generality, we choose the coordinate system $\vec{k}=(0,\,0,\, k)$
and parameterize $\vec{k}'=k\left(\cos\varphi\sin\vartheta,\,\sin\varphi\sin\vartheta,\,\cos\vartheta\right)$
since $k'=k$. We set $\vec{a}=\left(\vec{E}\,\fExpCoeffZero+\vec{P}\times\vec{E}\, D_{k}\right)$
and Eq.~(\ref{eq:Icoll}) is obtained from \begin{align}
 & \int d\Omega(\vec{k}')\: I\left(\vartheta\right)\,\left(\vec{k}-\vec{k}'\right)\cdot\vec{a}\nonumber \\
 & =\int d\Omega\: I\left(\vartheta\right)\,\big[k\, a_{z}\left(1-\cos\vartheta\right)-k\, a_{x}\cos\varphi\sin\vartheta\nonumber \\
 & \qquad-k\, a_{y}\sin\varphi\sin\vartheta\big]\nonumber \\
 & =\vec{k}\cdot\vec{a}\:\int d\Omega\: I\left(\vartheta\right)\,\left(1-\cos\vartheta\right),\end{align}
where we used in the last line that integration over $d\Omega$ removes
the terms in $\sin\varphi$ and $\cos\varphi$ and that $k\, a_{z}=\vec{k}\cdot\vec{a}$.
Next, we note that $\vec{n}=\vec{k}'\times\vec{k}/\left|\vec{k}'\times\vec{k}\right|=\left(\sin\phi,\,-\cos\phi,\,0\right)$
and insert into the left hand side of Eq.~(\ref{eq:IScoll}),\begin{align}
 & \int d\Omega(\vec{k}')\: I\left(\vartheta\right)S\left(\vartheta\right)\left(\bsigma\cdot\vec{n}\right)\left(\vec{k}\cdot\vec{E}-\vec{k}'\cdot\vec{E}\right)\fExpCoeffZero\nonumber \\
 & =\int d\Omega\left(\vec{k}'\right)\: I\left(\vartheta\right)S(\vartheta)\,\left(\sigma_{x}\sin\varphi-\sigma_{y}\cos\varphi\right)\nonumber \\
 & \qquad\times\big[k\, E_{z}\left(1-\cos\vartheta\right)-k\, E_{x}\cos\varphi\sin\vartheta\nonumber \\
 & \qquad\quad-k\, E_{y}\sin\varphi\sin\vartheta\big]C_{k}\nonumber \\
 & =-\frac{1}{2}\left(k\,\sigma_{x}E_{y}-k\,\sigma_{y}E_{x}\right)C_{k}\int d\Omega\left(\vec{k}'\right)\: I\left(\vartheta\right)S(\vartheta)\,\sin\vartheta,\end{align}
using that also terms in $\sin\varphi\cos\varphi$ vanish after integration
and that $\int d\varphi\,\cos^{2}\varphi=\int d\varphi\,\sin^{2}\varphi=\frac{1}{2}\int d\varphi$.
From $k\,\sigma_{x}E_{y}-k\,\sigma_{y}E_{x}=\vec{k}\cdot(\bsigma\times\vec{E})$,
we recover the r.h.s of Eq.~(\ref{eq:IScoll}).


\begin{thebibliography}{10}
\bibitem{KatoSpinHall}Y.K. Kato, R.C. Myers, A.C. Gossard, and D.D. Awschalom, Science \textbf{306},
1910 (2004). 
\bibitem{Wunderlich05}J. Wunderlich, B. Kaestner,  J. Sinova, and T. Jungwirth, Phys. Rev.
Lett. \textbf{94}, 047204 (2005). 
\bibitem{DP71}M.I. Dyakonov and V.I. Perel, Phys. Lett. \textbf{35A}, 459 (1971). 
\bibitem{Hirsch99}J.E. Hirsch, Phys. Rev. Lett. \textbf{83}, 1834 (1999). 
\bibitem{Zhang00}S. Zhang, Phys. Rev. Lett. \textbf{85}, 393 (2000). 
\bibitem{MottMassey}N.F. Mott and H.S.W. Massey, \emph{The Theory of Atomic Collisions}
(Oxford University Press, 1965). 
\bibitem{Murakami03}S. Murakami, N. Nagaosa, and S.-C. Zhang, Science \textbf{301}, 1348
(2003). 
\bibitem{Sinova04}J. Sinova, D. Culcer, Q. Niu, N.A. Sinitsyn, T. Jungwirth, and A.H.
MacDonald, Phys. Rev. Lett. \textbf{92}, 126603 (2004). 
\bibitem{Nozieres73}P. Nozi\`eres and C Lewiner, J. Phys. (Paris) \textbf{34}, 901 (1973),
their SO coupling constant is $\lambda_{\mathrm{NL}}=2\lambda/\hbar$. 
\bibitem{CrepieuxBrunoAHE}A. Cr\'epieux and P. Bruno, Phys. Rev. B \textbf{64}, 014416 (2001). 
\bibitem{SHList}See, e.g., A.A. Burkov, A.S. N\'u\~nez, and A.H. MacDonald, Phys.
Rev. B \textbf{70}, 155308 (2004); S.I. Erlingsson, J. Schliemann,
and D. Loss, \emph{ibid.} \textbf{71}, 035319 (2005); O.V. Dimitrova,
cond-mat/0405339v2. 
\bibitem{Mish04}E.G. Mishchenko, A.V. Shytov, and B.I. Halperin, Phys. Rev. Lett. \textbf{93},
226602 (2004). 
\bibitem{fnKatoModel}In Ref.~\onlinecite{KatoSpinHall}, a model with spin diffusion and
spin relaxation is used to relate observed spin accumulation at the
boundaries to a bulk spin Hall current.
\bibitem{R03}E.I. Rashba, Phys. Rev. B \textbf{68}, 241315 (2003). 
\bibitem{Dresselhaus55}G.Dresselhaus, Phys. Rev. \textbf{100}, 580 (1955). 
\bibitem{BernevigZhangIntrinsic3D}B.A. Bernevig and S.-C. Zhang, cond-mat/0412550. 
\bibitem{Usaj04}G. Usaj and C.A. Balseiro, cond-mat/0405065v2. 
\bibitem{Nikolic2DBoundary}B.K. Nikolic, S. Souma, L.P. Zarbo, and J. Sinova, cond-mat/0412595. 
\bibitem{KatoAwschalomPrivate}Y.K. Kato and D.D. Awschalom (private communication).
\bibitem{Winkler}R. Winkler, \emph{Spin-Orbit Coupling Effects in Two-Dimensional Electron
and Hole System} (Springer, 2003). 
\bibitem{BergerSideJump}L. Berger, Phys. Rev. B \textbf{2}, 4559 (1970). 
\bibitem{MotzRMP}J.W. Motz, H. Olsen, and H.W. Koch, Rev. Mod. Phys. \textbf{36}, 881
(1964), ${}$Sec.\ VI and Eqs.\ (1A-107) and (1A-403). 
\bibitem{VoskoboynikovPRB}H.C. Huang, O. Voskoboynikov, and C.P. Lee, Phys. Rev. B \textbf{67},
195337 (2003). 
\bibitem{Skewness2D}The skew-scattering contribution for a 2D system is obtained by replacing
$\frac{1}{2}\ShermanBmAvg\to\ShermanBmAvg^{2\mathrm{D}}$ in Eqs.~(\ref{eq:f})--(\ref{eq:SS}),
where $\ShermanBmAvg^{2\mathrm{D}}$ is given by Eq.~(\ref{eq:GammaSkew})
after replacing $\int d\Omega\to\int_{0}^{\pi}d\vartheta$. For example,
using the same parameters as in the text, we obtain $\ShermanBmAvg^{2\mathrm{D}}\approx1/1300$.
However, in 2D, both transport and calculation of cross sections \cite{VoskoboynikovPRB}
are complicated by the presence of intrinsic $k$-linear terms (Rashba
and Dresselhaus) which are difficult to avoid. 
\bibitem{Elliott}R.J. Elliott, Phys. Rev. \textbf{96} (1954). 
\bibitem{YafetSSP}Y. Yafet, Solid State Physics \textbf{14}, 1 (1963), eds. F. Seitz
and D. Turnbull (Academic, NY). In modern terms, $\delta\dot{\vec{r}}_{\mathrm{SO}}$
is related to the Berry curvature.
\bibitem{fnRescale}We do not renormalize higher-order or spin-independent relativistic
terms; their effect should be small and is disregarded.
\bibitem{fnkFell}From $\sigma_{xx}$ we extract a rather small value $k_{\mathrm{F}}\ell\approx4$,
i.e., the accuracy of the Boltzmann approach is restricted.
\bibitem{fnQualitative}Qualitatively, our model also agrees with the main observations. First,
because Eq.~(\ref{eq:H}) is based on $s$- and $p$-band interaction,
our model is isotropic, consistent with the observation of an isotropic
spin Hall effect.  Second, it is compatible with the interpretation
of the shifted Kerr rotation data due to strain-induced spin-orbit
fields, see Fig.~4 of Ref.~\onlinecite{KatoSpinHall}.
\bibitem{Inoue04}J.I. Inoue, G.E.W. Bauer, and L.W. Molenkamp, Phys. Rev. B \textbf{70},
041303 (2004). 
\bibitem{Chalaev2D}O. Chalaev and D. Loss, cond-mat/0407342v2. 
\bibitem{MishenkoPrivate}E.G. Mishchenko (private communication); S.-C. Zhang (private communication).
\end{thebibliography}
\end{document}